\documentstyle[12pt,epsfig]{article}

\def\d{{\rm d}}
\def\qup{q_{\uparrow}}
\def\qdown{q_{\downarrow}}

\begin{document}
\sloppy
\thispagestyle{empty}

\begin{flushright}
DTP/95/78   \\
September 1995 \\
\end{flushright}
\vskip 1.cm
\mbox{}
\vspace*{\fill}
\begin{center}
{\LARGE\bf Present Status of Polarized } \\

\vspace{2mm}
{\LARGE\bf Parton Distributions}\\

\vspace{2em}
\large
T. Gehrmann and W.J. Stirling
\\
\vspace{2em}
{\it Department of Physics, University of Durham, \\
 Durham~DH1~3LE, England}\\
\end{center}
\vspace*{\fill}
\begin{abstract}
\noindent
A review of the present knowledge on polarized parton distributions is
given. The effects of perturbative evolution on these distributions
are discussed qualitatively and a comparison of various recent
parametrizations is made.
\end{abstract}
\vspace*{\fill}
\begin{center}
{\it Talk given at the
Workshop on the Prospects of Spin Physics at HERA,
DESY-Zeuthen, August 28-31, 1995}
\end{center}
\vspace*{\fill}
\newpage
%
\section{Introduction}
\label{sect1}

A series of precision measurements~\cite{newexp} of the polarized
structure function
$g_1(x,Q^2)$ off proton and deuteron targets has considerably improved
our knowledge on the spin structure of the nucleon over the last two
years. In combination with several older measurements~\cite{oldexp}, these
experiments now cover an $x$-range of $0.003\leq x \leq 0.8$, although
the $Q^2$ range for fixed $x$ is still rather restricted.

In the `naive' parton model  $g_1$ can, like the
unpolarized structure function $F_1$, be expressed in terms of the
probability distributions for finding quarks with spin parallel or
antiparallel to the longitudinally polarized parent proton:
\begin{eqnarray}
F_1(x,Q^2) &=& \frac{1}{2} \sum_q\; e_q^2\;  [q(x) +\bar q(x) ]  \\
g_1(x,Q^2) &=& \frac{1}{2} \sum_q\; e_q^2\;  [\Delta q(x) +\Delta\bar q(x) ]\;
,
\label{naiveg1}
\end{eqnarray}
where
\begin{equation}
q = \qup + \qdown\; , \quad \Delta q = \qup-\qdown \; .
\end{equation}
Furthermore, the naive parton model predicts exact scaling behaviour
for the above distributions, i.e. independence of the $Q^2$ scale of
the measurement. These distributions are intrinsic, nonperturbative
features of the nucleon, and can therefore
at present  only be determined from a fit
to the structure function data.
Some insight can however be gained from thermodynamical
models~\cite{soffer} or from the light-cone wavefunctions of partons
in the nucleon~\cite{bro2}.

Perturbative QCD yields corrections~\cite{ap} to the simple parton model
picture, which are manifest in a scale-dependence of the parton
distributions. The quantitative features of these corrections will be
discussed in Section~2.

Various groups have used the recent data on $g_1(x,Q^2)$ to determine
the polarized parton distributions, taking into account the
leading-order QCD corrections\footnote{The next-to-leading order QCD
corrections to the scale dependence of polarized parton distributions
have only been calculated very recently~\cite{vn}. So far, only one
group has used these to produce a set of NLO polarized distributions
\cite{grvnlo}.}. The concepts of the various
approaches and their results are compared in section 3. Finally,
section 4 contains a brief summary.

\section{Evolution of polarized parton distributions}
\label{sect2}

In the QCD corrected parton model $g_1$ is expressed in terms
of  parton densities
for the polarization of quarks and gluons,
\begin{eqnarray}
g_1(x,Q^2) &=& {1\over 2}\; \sum_q\; e_q^2 \; \int_x^1 {dy\over y}
\;  \left[\Delta q(x/y, Q^2) +\Delta\bar q(x/y,Q^2) \right]\; \nonumber \\
& & \times  \left\{\delta(1-y)
+ {\alpha_s(Q^2) \over 2\pi} \Delta C_q(y) + \ldots \right\} \nonumber \\
& & + \; \frac{1}{9} \int_x^1 \; {dy\over y}\;  \Delta G(x/y,Q^2) \; \left\{
n_f\;   {\alpha_s(Q^2) \over 2\pi}\;\Delta C_G(y) +\ldots \right\}.
\label{qcdg1}
\end{eqnarray}

The emission of collinear partons
gives rise to an evolution of the parton densities \cite{ap},
\begin{eqnarray}
\label{evo}
\lefteqn{\frac{\partial}{\partial \ln Q^2}
\left(\begin{array}{c} \Delta q \\ \Delta G
\end{array} \right) (x,Q^2)} \nonumber \\
& & =
\frac{\alpha_s(Q^2)}{2\pi}
\int_x^1 \frac{\d y}{y} \left(\begin{array}{cc} \Delta P_{qq}  & \Delta
P_{qg} \\ 2 n_f \Delta P_{gq} & \Delta P_{gg}\end{array} \right) (y)\;
\left(\begin{array}{c} \Delta
q \\ \Delta G \end{array} \right)(x/y,Q^2).
\end{eqnarray}
This equation only determines how the
distributions change with $Q^2$, not the distributions themselves.
The boundary conditions for the solution enter as initial distributions
$\Delta q(x,Q_0^2)$ and $\Delta G(x,Q_0^2)$.

Even if the precise form of the polarized parton distributions is still
not yet known, several qualitative features of the $Q^2$
dependence can be determined from the splitting functions (Fig.~1).
The evolution of the polarized valence quark density is identical to
the unpolarized one: the distribution decreases in the large $x$
region and increases for smaller $x$. The polarized singlet
distribution is changed by two splitting processes: $q\rightarrow q$
lowers the distribution at large $x$ and increases at  small
$x$ while $g\rightarrow q$ slightly increases it at large $x$ and
lowers it at small $x$. The overall change of $\Delta \Sigma
(x,Q^2)$ is therefore sensitive to the relative magnitude of the quark
and gluon polari- zations. The evolution of $\Delta G(x,Q^2)$ is
dominated by the $g\rightarrow g$ splitting, which strongly increases
the distribution at small and medium $x$.
Only if the gluon polarization is initially smaller
than the total quark polarization, will effects from $q\rightarrow g$
contribute visibly to the increase of  $\Delta G(x,Q^2)$.

\begin{center}
{}~ \epsfig{file=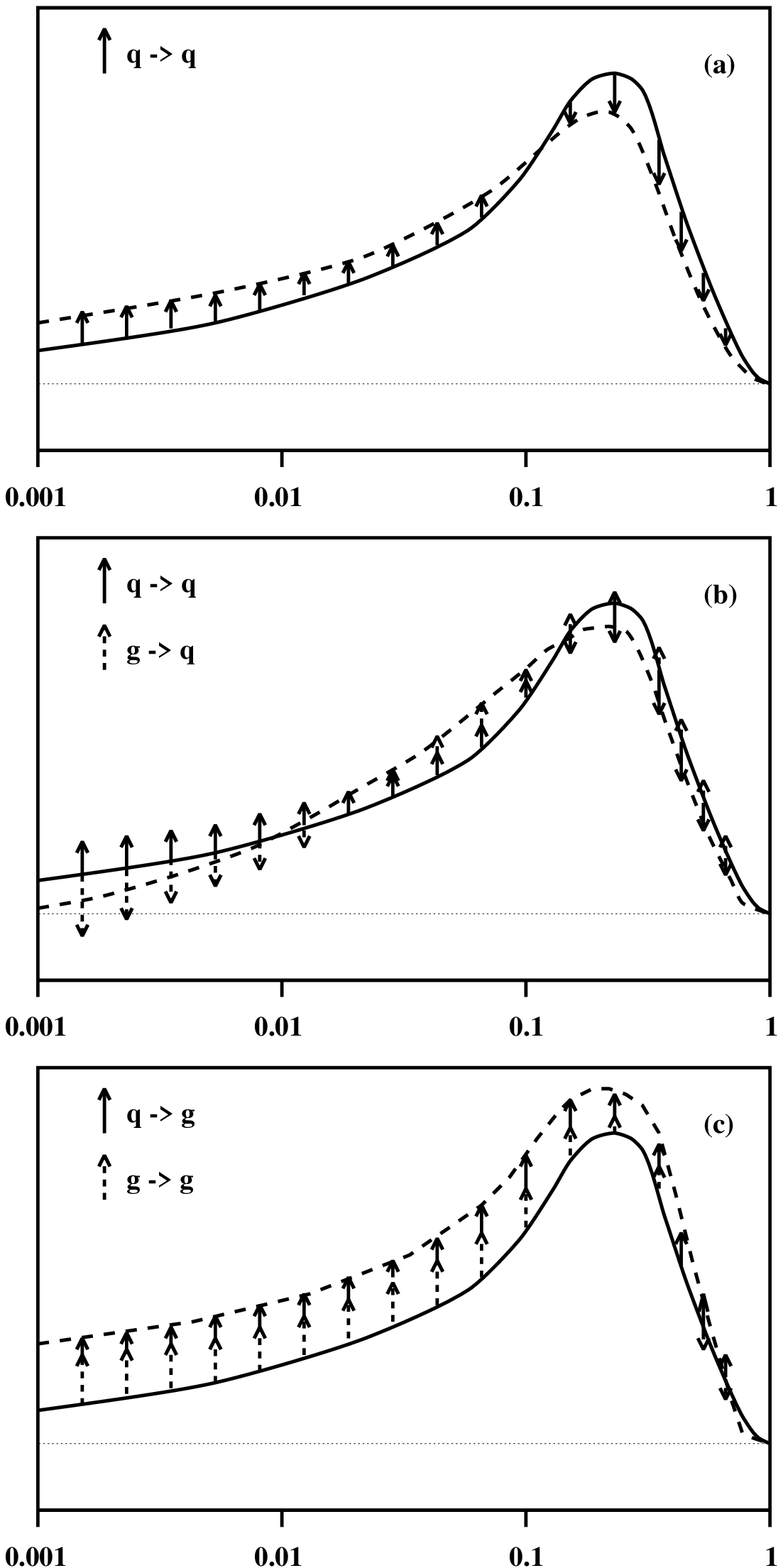,height=16.9cm}
\small
\end{center}
{\sf Figure~1:}~Qualitative description of the evolution of polarized
valence quark~(a), polarized singlet quark~(b) and polarized gluon~(c)
densities.  Solid line: $x\ast f(x)$ at $Q_0^2$, dashed line: $x\ast
f(x)$ at $Q^2 > Q_0^2$.
\normalsize

\section{Parametrizations of polarized parton distributions}
\label{sect3}

Although the change of the polarized parton distributions with
increasing $Q^2$ is determined by the above evolution equations, the
distributions themselves are incalculable in perturbative QCD. The
starting distributions at some low scale $Q_0^2$ reflect the {\it
nonperturbative} spin-structure of the nucleon and can only be fitted
to the experimental data. Various groups have performed these fits
using the leading-order evolution equations\footnote{While this review
was in preparation, the first set of next-to-leading order polarized
parton distributions has appeared \protect{\cite{grvnlo}}.}.
In the following we will restrict our discussion to the most
recent parametrizations from Gl\"uck-Reya-Vogelsang
(GRV \cite{grvpol}), the La Plata group (LP \cite{lp}) and to our own results
(GS \cite{gs94,gs95}). These parametrizations differ in various
aspects:
\begin{itemize}
\item[{(i)}] {\bf Data selection:}
\begin{table}[htb]
\begin{center}
\begin{tabular}{|l|c|c|c|c|c|c|}\hline
\rule[-1.2ex]{0mm}{4ex}expt. & $x$ & $Q^2[\mbox{GeV}^2]$ & GRV & LP & GS94 &
GS95 \\ \hline
\rule[-1.2ex]{0mm}{4ex} E130-p & 0.180 -- 0.7 & 3.5 -- 10 & & $\bullet$
& $\bullet$ & $\bullet$ \\ \hline
\rule[-1.2ex]{0mm}{4ex} EMC-p & 0.010 -- 0.7 & 1.5 -- 70 & $\bullet$ &
$\bullet$ & $\bullet$ & $\bullet$ \\ \hline
\rule[-1.2ex]{0mm}{4ex} SMC-p & 0.003 - 0.7 & 1.0 -- 60 & $\bullet$ &
$\bullet$ & $\bullet$ & $\bullet$ \\ \hline
\rule[-1.2ex]{0mm}{4ex} E143-p & 0.029 -- 0.8 & 1.3 -- 10 &  $\bullet$ &
&  & $\bullet$ \\ \hline
\rule[-1.2ex]{0mm}{4ex} SMC-d93 & 0.006 -- 0.6 & 1.0 -- 30 &  $\bullet$
& $\bullet$ & & $\bullet$ \\ \hline
\rule[-1.2ex]{0mm}{4ex} SMC-d95 & 0.003 -- 0.7 & 1.0 -- 60 & & & & $\bullet$
\\ \hline
\rule[-1.2ex]{0mm}{4ex} E143-d & 0.029 -- 0.8 & 1.0 -- 30 &  $\bullet$ &
&  & $\bullet$ \\ \hline
\rule[-1.2ex]{0mm}{4ex} E142-n & 0.030 -- 0.6 & 1.0 -- 10 &  $\bullet$ &
& $\bullet$ & $\bullet$ \\ \hline
\end{tabular}
\end{center}
\end{table}

\item[{(ii)}] {\bf Construction of $g_1(x,Q^2)$:} The above
experiments have measured the cross section asymmetry
\begin{equation}
A_1(x,Q^2) = \frac{g_1(x,Q^2)}{F_1(x,Q^2)}.
\end{equation}
Therefore, one has to use $F_1(x,Q^2)$ for the extraction of
$g_1(x,Q^2)$. The various parametrizations use either the unpolarized parton
distributions of Refs.~\cite{grvlo} (GRV) and \cite{mrsd} (LP) or the SLAC/NMC
parametrizations \cite{slac} (GS).

\item[{(iii)}] {\bf Positivity of the distributions:} The most
fundamental constraint on the polarized parton distribution is the
positivity of the individual  helicity distributions. This forces the
polarized distributions to be less in magnitude than the unpolarized
ones,
\begin{equation}
\mid\! \Delta f \! \mid \le f(x) \qquad \mbox{with } f=q,G .
\end{equation}
This constraint is either incorporated by defining polarizations
(GRV,LP)
\begin{equation}
\Delta f(x,Q_0^2) = \chi_f (x) f (x,Q_0^2) ,
\end{equation}
or by constraining combinations of parameters in the initial
distributions (GS).

\item[{(iv)}] {\bf Constraints from the Ellis-Jaffe sum rule:} The
Ellis-Jaffe sum rule \cite{ej}
\begin{equation}
\Gamma_1^{p,n} (Q^2) = \pm \frac{1}{12} a_3 + \frac{1}{36} a_8 +
\frac{1}{9} \Delta \Sigma^{(1)} (Q^2)
\end{equation}
relates the first moment of $g_1(x,Q^2)$ to the expectation values
\begin{eqnarray}
a_3 & = & \Delta u^{(1)}(Q^2) - \Delta d^{(1)}(Q^2) = F+D = 1.257 \nonumber \\
a_8 & = & \Delta u^{(1)}(Q^2) + \Delta d^{(1)}(Q^2) -2  \Delta
s^{(1)}(Q^2)  = 3F-D = 0.579 \nonumber \\
\end{eqnarray}
of the conserved nonsinglet axial vector currents. The nonconserved singlet
axial vector current
\begin{equation}
\Delta \Sigma^{(1)} (Q^2) = \Delta u^{(1)}(Q^2)  + \Delta d^{(1)}(Q^2)
+ \Delta s^{(1)} (Q^2)
\end{equation}
is then matched to the experimental data. Note that
the Ellis-Jaffe sum rule is independent of $Q^2$ in a leading-order
model. Therefore, only an average value of all experimental
measurements can be
reproduced. Furthermore, the above procedure is not unique, as
the different
parametrizations use different assumptions for the flavour
decomposition of the sea quark polarization.

\item[{(v)}] {\bf Constraints from asymptotic estimates:} The
experimental data on $g_1(x,Q^2)$ are presently insufficient for a
global determination of all polarized parton distributions. Therefore
some properties -- such as the behaviour at small and large $x$ --
have to be estimated from theoretical considerations. The estimates used
differ from parametrization to parametrization, the most common being:
color coherence and Regge arguments at small $x$,
and counting rules at large $x$ (see Ref.~\cite{bro2} for a recent review).
\end{itemize}

As the data on $g_1^p$ and $g_1^d$ have almost the same statistical
quality, the non-singlet distributions $\Delta u_{val}(x,Q_0^2)$ and
$\Delta d_{val}(x,Q^2_0)$ can be determined from these measurements.
Figure 2 illustrates the agreement between the
parametrizations. The differences arise mainly from the data included
in the fit and from different assumptions on the magnitude of the sea
quark polarization entering the Ellis-Jaffe sum rule.

\begin{center}
\mbox{\epsfig{file=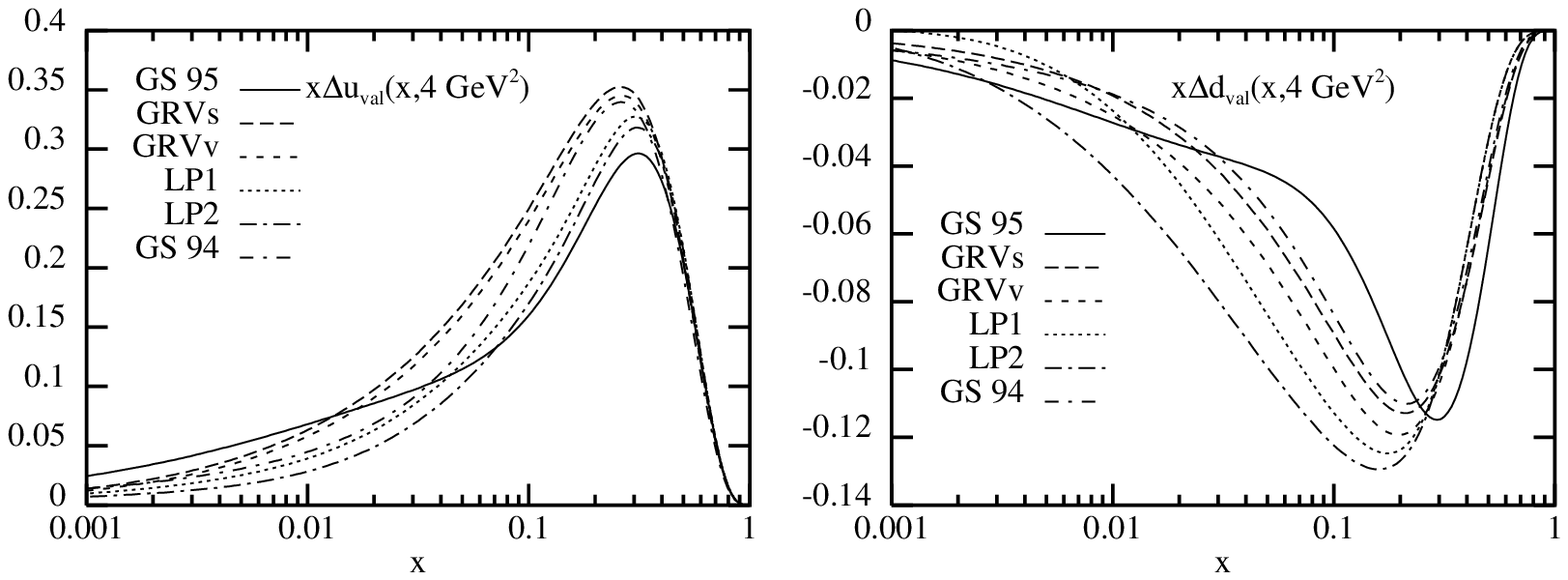,height=5cm}}
\small
\end{center}
{\sf Figure~2:}~Parametrizations of the polarized valence quark
distributions.
\vspace{3mm}
\normalsize

The sea quark and gluon distributions are much more sensitive to the
various constraints used to determine the distributions. As these
differ for all parametrizations, the corresponding distributions span a wide
range
of possibilities, as can be seen in Figs.~3 and 4.

\begin{center}
\mbox{\epsfig{file=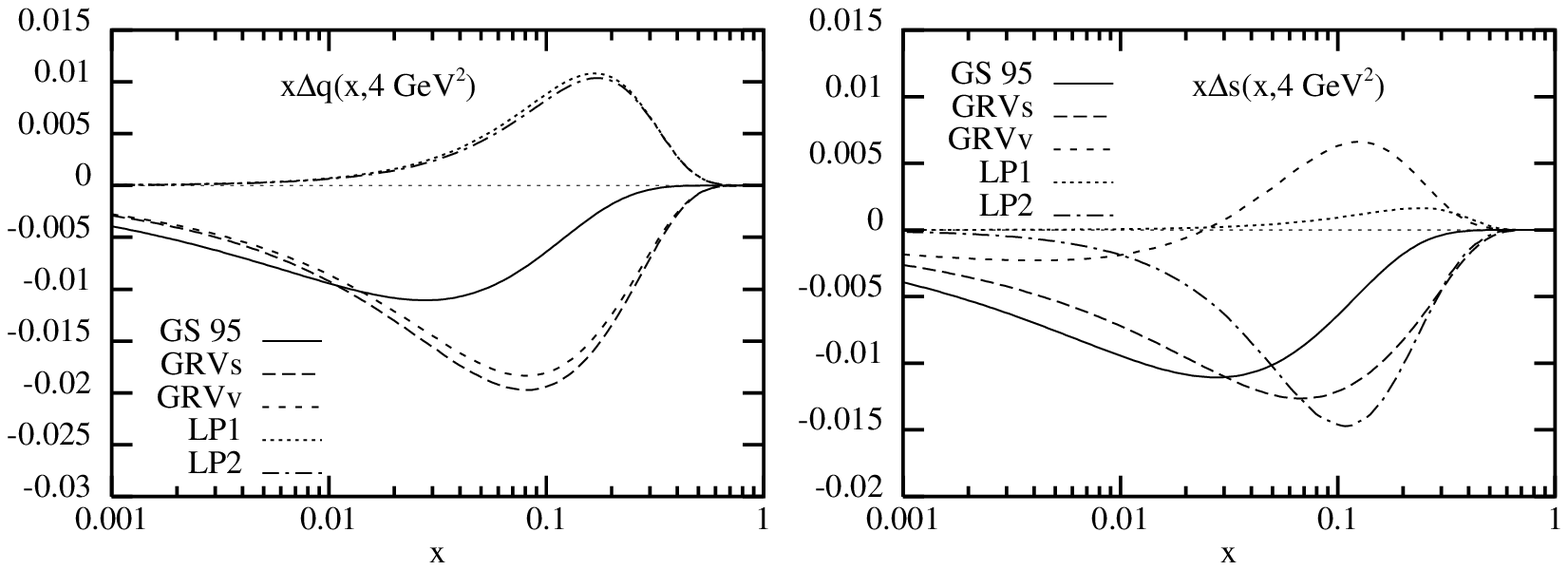,height=5cm}}
\small
\end{center}
{\sf Figure~3:}~Parametrizations of the polarized sea
($q=u=\bar{u}=d=\bar{d}$) and polarized strange ($s=\bar{s}$)
quark distributions.
\vspace{3mm}
\normalsize

\begin{center}
\mbox{\epsfig{file=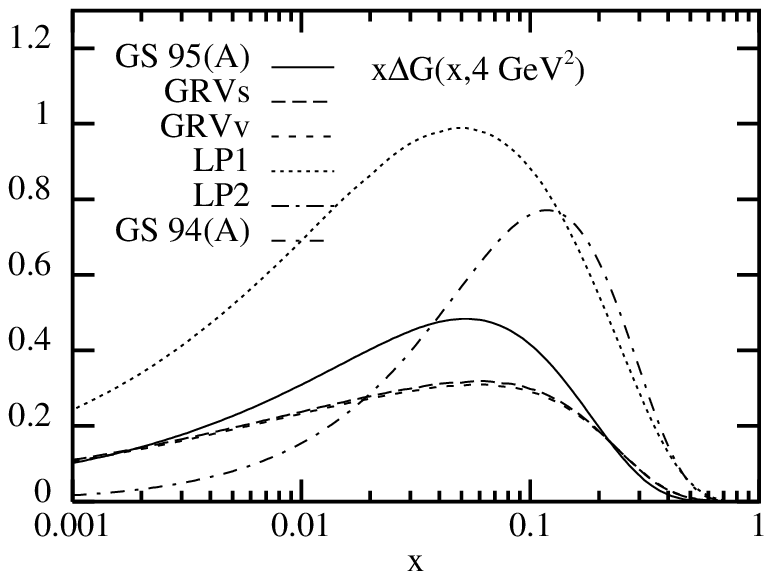,height=5cm}}
\small
\end{center}
{\sf Figure~4:}~Parametrizations of the polarized gluon distribution.
\vspace{3mm}
\normalsize

Without experimental data on processes other than the structure
function $g_1$, it will be very difficult to distinguish between
these different possiblities.
Future experiments at RHIC,
CERN and SLAC, together with final state measurements at a polarized
HERA collider \cite{futexp} and at HERMES, could provide this
information on $\Delta \bar{q}(x)$ and $\Delta G(x)$ from various
different processes.

\section{Conclusions}
\label{sect4}
In this talk, we have given a brief review of the quantitative
features of the $Q^2$ evolution of polarized parton densities.
Comparing various recent parametrizations of these densities, we have
found that the present data on $g_1^{p,n}(x,Q^2)$ determine the
valence quark polarization in the nucleon to some accuracy. In
contrast, the polarization of the quark sea and the gluon are strongly
dependent on additional theoretical constraints imposed on the
distributions. This situation is not expected to improve significantly with
more
precise data on $g_1$, which
clearly shows the need for complementary
measurements on polarized nucleons.

\section*{Acknowledgements}

\noindent  Financial support from  the UK PPARC (WJS), and from
the Gottlieb Daimler- und Karl Benz-Stiftung and the
Studienstiftung des deutschen Volkes (TG) is gratefully acknowledged.
This work was supported in part by the EU Programme
``Human Capital and Mobility'', Network ``Physics at High Energy
Colliders'', contract CHRX-CT93-0357 (DG 12 COMA).
\goodbreak


\newpage

\end{document}